\documentclass[preprint,aps,showpacs]{revtex4-1}
\usepackage{amsmath,amssymb,amsfonts,dcolumn,color,graphicx,graphics,latexsym,placeins,epsfig,tikz}
\usetikzlibrary{arrows,shapes}
\usepackage{subfigure,rotating,hyperref,bm}

\newcommand{\be}{\begin{equation}}
\newcommand{\ee}{\end{equation}}
\newcommand{\ba}{\begin{eqnarray}}
\newcommand{\ea}{\end{eqnarray}}

\begin{document}

\title{Wormholes with nonexotic matter in Born-Infeld gravity}

\author{Rajibul Shaikh}
\email{rajibul.shaikh@tifr.res.in}
\affiliation{Tata Institute of Fundamental Research, Homi Bhabha Road, Colaba, Mumbai 400005, India.}

\begin{abstract}
We show that, in contrast to general relativity, in Eddington-inspired Born-Infeld gravity (EiBI), a violation of the null convergence condition does not necessarily lead to a violation of the null energy condition, by establishing a relationship between them. This serves as a motivation for finding wormhole solutions which can be supported by nonexotic matter in this gravity theory. We then obtain exact solutions of the field equations in EiBI gravity coupled to arbitrary nonlinear electrodynamics and anisotropic fluids. Depending on the signs and values of different parameters, the general solutions can represent both black holes and wormholes. In this work, we analyze the wormhole solutions. These wormholes are supported by nonexotic matter, i.e., matter satisfying all the energy conditions. As special cases of our general solutions, we work out several specific examples by considering Maxwell, power-law, Born-Infeld electrodynamics models and a particular form of an anisotropic fluid.

\end{abstract}


\maketitle

\section{Introduction}
Wormholes are theoretical constructs constituting short cuts or tunnels or openings to otherwise distant parts of the cosmos or different universes \citep{morris1,visser1}. The idea of such spacetime geometries came from Einstein and Rosen's proposal of the Einstein-Rosen bridge \citep{ER_bridge}. The term {\em wormhole} was coined by Misner and Wheeler \citep{wormhole_first}. One of the first wormhole solutions, the Ellis-Bronnikov wormhole, was found in the framework of general relativity (GR), using a wrong sign (phantom) in the scalar field Lagrangian \citep{Ellis,Bronnikov}. Subsequently, the possibility of having time-machine models using wormholes was introduced \citep{time_machine_1,time_machine_2,time_machine_3}. This led to growing interest in wormholes \citep{visser1}.

A wormhole can be thought of as a defocusing lens in the sense that an initially converging family of radial null rays, while passing through the wormhole, first becomes parallel at the wormhole throat and then starts diverging on the other side. This defocusing of the family of null rays passing through the wormhole is the outcome of the fact that the null convergence condition (NCC) is violated in the vicinity or at least at the wormhole throat \citep{visser1}, as can be seen from an analysis of the Raychaudhuri equation for this family. In GR, a violation of the NCC leads to a violation of the null energy condition (NEC) which, in turn, leads to violations of the other various energy conditions (weak, strong, dominant, etc.) \cite{hawking,wald}. Therefore, in the framework of GR, wormholes require energy-condition-violating matter (often termed ``exotic matter") to be supported \cite{morris1,visser1}.

However, the above-mentioned requirement of exotic matter to support a wormhole can be avoided in many alternative or modified theories of gravity. In such gravity theories, since the structures of the field equations are different from that of GR, the violation of the NCC does not necessarily lead to the violations of the various energy conditions \cite{capozziello1,capozziello2}. Therefore, in such theories, wormholes can be supported by matter which satisfies all the energy conditions but violates the convergence condition. See \cite{bhawal_1992,maeda_2008,lobo_2009,dehghani_2009,kanti_2011,garcia_2011,boemer_2012,
harko_2013,kar_2015,bronnikov_2015,bambi_2016,rajibul_2016,myrzakulov_2016,moraes_2017,moradpour_2017,
hohmann_unpub,ovgun_unpub} and references therein for some such examples. Also see \cite{nandi_2006,tsukamoto_2012,rajibul_2017,jusufi_2018a,jusufi_2018b,bambi_2013a,nedkova_2013,
ohgami_2015,abdujabbarov_2016b,rajibul_2018b,amir_2018,harko_2009,bambi_2013b,cardoso_2016,
konoplya_2016,aneesh_2018} for some recent works on different aspects of wormholes.

Eddington inspired Born-Infeld gravity (EiBI) \citep{banados}, which is a modified theory of gravity, belongs to a class of Born-Infeld inspired gravity theory first proposed by Deser and Gibbons \citep{deser}, inspired by the earlier work of Eddington \citep{eddington} and the nonlinear electrodynamics of Born and Infeld \citep{born}. The theory is equivalent to Einstein's GR in vacuum but differs from it within matter. Since its introduction, various aspects of EiBI gravity have been studied by many researchers in the recent past, including black holes \citep{banados,BH_1,BH_2,karan_2015,BH_3,soumya_2015,BH_5,BH_6,BH_7,jana_shaikh,BH_8}, wormholes \citep{WH_1,rajibul_2015,WH_2,WH_3}, compact stars \citep{CS_1,CS_2,CS_3,CS_4,CS_5}, cosmological aspects \citep{banados,delsate1,cho1,COS_1,COS_2,perturbation1,perturbation2,COS_3,COS_4,COS_5,COS_6,COS_7}, astrophysical aspects \citep{ASTRO_1,ASTRO_2,ASTRO_3}, gravitational collapse \citep{collapse_1,rajibul_2018a}, gravitational waves \citep{GW_1,GW_2}, implications in nongravitational contexts like particle physics \citep{EiBI_particle} etc. See \cite{EiBIreview} for a recent review on various studies in EiBI gravity. In this work, we first show that, in this gravity theory, the violation of the NCC does not necessarily lead to the violation of the NEC. We then obtain exact solutions of the field equations in EiBI gravity coupled to arbitrary nonlinear electrodynamics and anisotropic fluids. The general solutions can represent both black holes and wormholes. In this work, we focus on the wormhole solutions which are supported by nonexotic matter.

The plan of the paper is as follows. In the next section, we briefly recall the EiBI theory. In Sec. \ref{sec:NCC}, we establish a relationship between the NCC and the NEC along a congruence of radial null geodesics. In Sec. \ref{sec:wormholes}, we obtain exact solutions, which represent both black holes and wormholes, of the field equations in EiBI gravity coupled to arbitrary nonlinear electrodynamics and anisotropic fluids and analyze the wormhole solutions. We work out some specific examples in Sec. \ref{sec:examples}. We conclude in Sec. \ref{sec:conclusion} with a summary of the key results.

\section{Eddington-inspired Born-Infeld gravity}
\label{sec:EiBI}

The action in EiBI gravity developed in \cite{banados} is given by
\begin{eqnarray}
S_{BI}[g,\Gamma,\Psi]&=&\frac{c^4}{8\pi G\kappa}\int d^4x\left[\sqrt{-\left\vert g_{\mu\nu}+\kappa R_{\mu\nu}(\Gamma)\right\vert}-\lambda \sqrt{-g}\right]+S_{M}(g,\Psi),
\nonumber
\end{eqnarray}
where $c$ is the speed of light, $G$ is Newton's gravitational constant, $\lambda=1+\kappa\Lambda$, $\kappa$ is the EiBI theory parameter, $\Lambda$ is the cosmological constant, $R_{\mu\nu}(\Gamma)$ is the symmetric part of the Ricci tensor built with the independent connection $\Gamma$,  $S_{M}(g,\Psi)$ is the action for the matter field, and the vertical bars stand for the matrix determinant. Variations of this action with respect to the metric tensor $g_{\mu\nu}$ and  the connection $\Gamma$ yield, respectively, \citep{banados,cho1,delsate1}
\begin{equation}
\sqrt{-q}q^{\mu\nu}=\lambda \sqrt{-g}g^{\mu\nu}-\bar{\kappa} \sqrt{-g}T^{\mu\nu}
\label{eq:field_equation1}
\end{equation}
\begin{equation}
\nabla^\Gamma_\alpha \left(\sqrt{-q} q^{\mu\nu}  \right)=0,
\label{eq:metric_compatibility}
\end{equation}
where $\bar{\kappa}=\frac{8\pi G\kappa}{c^4}$, $\nabla^\Gamma$ denotes the covariant derivative defined by the connection $\Gamma$ and $q^{\mu\nu}$ is the inverse of the auxiliary metric $q_{\mu\nu}$ defined by
\begin{equation}
q_{\mu\nu}=g_{\mu\nu}+\kappa R_{\mu\nu}(\Gamma).
\label{eq:field_equation2}
\end{equation}
In obtaining the field equations from the variation of the action, it is assumed that both the connection $\Gamma$ and the Ricci tensor $R_{\mu\nu}(\Gamma)$ are symmetric, i.e., $\Gamma^\mu_{\nu\rho}=\Gamma^\mu_{\rho\nu}$ and $R_{\mu\nu}(\Gamma)=R_{\nu\mu}(\Gamma)$. Equation (\ref{eq:metric_compatibility}) gives the metric compatibility equation which yields
\begin{equation}
\Gamma^\mu_{\nu\rho}=\frac{1}{2}q^{\mu\sigma}\left(q_{\nu\sigma,\rho}+q_{\rho\sigma,\nu}-q_{\nu\rho,\sigma} \right).
\end{equation}
Therefore, the connection $\Gamma^\mu_{\nu\rho}$ is the Levi-Civita connection of the auxiliary metric $q_{\mu\nu}$. Either in vacuum or in the limit $\kappa\to 0$, GR is recovered \citep{banados}.

\section{Convergence condition and energy conditions in EiBI gravity}
\label{sec:NCC}

As we have mentioned in the Introduction, at or in the vicinity of a wormhole throat, the NCC is violated along a congruence of radial null geodesics passing through it, and unlike in GR, in many modified gravity theories, the violation of the NCC may not lead to violations of the different energy conditions. In this section, we explore this possibility in the context of EiBI gravity. To study the NCC along a radial null geodesic congruence, we consider, respectively, the following $Ans\ddot{a}tze$ for the physical and the auxiliary metric:
\begin{equation}
ds_g^2=-e^{2\alpha(r)} dt^2+e^{2\beta(r)} dr^2+r^2(d\theta ^2+\sin ^2\theta d\phi ^2),
\label{eq:physical_metric0}
\end{equation}
\begin{equation}
ds_q^2=-e^{2\nu(r)} dt^2+e^{2\Psi(r)} dr^2+H^2(r)(d\theta ^2+\sin ^2\theta d\phi ^2).
\label{eq:auxiliary_metric0}
\end{equation}
For an energy-momentum tensor of the form $T^\mu_{\;\nu}={\rm diag}(-\rho,p_r,p_\theta,p_\theta)$, the field equation (\ref{eq:field_equation1}) yields
\begin{equation}
e^{\alpha(r)}=e^{\nu(r)}\sqrt{\tau(1+\kappa\rho)}, \hspace{0.3cm} e^{\beta(r)}=e^{\Psi(r)}\sqrt{\tau(1-\kappa p_r)} , \hspace{0.3cm} r=H(r)\sqrt{\tau(1-\kappa p_{\theta})},
\label{eq:relations}
\end{equation}
where $\tau=\frac{1}{\sqrt{(1+\kappa\rho)(1-\kappa p_r)(1-\kappa p_{\theta})^2}}$. Here, we have taken $G=c=1$, $\Lambda=0$ and $8\pi=1$. Later, in the matter Lagrangian also, we will set $8\pi=1$. This is for convenience. Note that the Ricci tensor appearing in the field equation (\ref{eq:field_equation2}) is that of the auxiliary metric $q_{\mu\nu}$. But, the Ricci tensor appearing in the NCC is that of the physical metric $g_{\mu\nu}$. However, for a family of radial null geodesics with four-velocity $k^{\alpha}$, we can express the NCC in terms of $\rho$, $p_r$, $p_\theta$ and their derivatives by using (\ref{eq:field_equation2}) and (\ref{eq:relations}). To this end, we first note that, for a family of radial null geodesics in the equatorial plane of the physical metric (\ref{eq:physical_metric0}), $k^t=e^{-2\alpha}$ and $k^r=\pm e^{-(\alpha+\beta)}$. Therefore, using (\ref{eq:relations}), we obtain
\begin{eqnarray}
R_{\mu\nu}(\Gamma)k^\mu k^\nu &=& e^{-2\alpha}\left[-R^t_t(\Gamma)e^{2(\nu-\alpha)}+R^r_r(\Gamma) e^{2(\Psi-\beta)}\right] \nonumber\\
&=& \frac{\kappa(\rho+p_r)e^{-2\alpha}}{\tau(1+\kappa\rho)(1-\kappa p_r)}R^t_t(\Gamma)+\frac{e^{-2\alpha}}{\tau(1-\kappa p_r)}\left[-R^t_t(\Gamma)+R^r_r(\Gamma)\right],
\label{eq:cc1}
\end{eqnarray}
where $R_{\mu\nu}(\Gamma)$ is the Ricci tensor of the auxiliary metric $q_{\mu\nu}$ and its indices are raised by using the auxiliary metric (\ref{eq:auxiliary_metric0}). Using the field equation $\kappa R_{\mu\nu}(\Gamma)=q_{\mu\nu}-g_{\mu\nu}$, the null geodesic equation $g_{\mu\nu}k^\mu k^\nu=0$ and Eqs. (\ref{eq:physical_metric0})-(\ref{eq:relations}), we obtain, along the family of radial null geodesics,
\begin{equation}
R_{\mu\nu}(\Gamma)k^\mu k^\nu=\frac{1}{\kappa}(q_{\mu\nu}-g_{\mu\nu})k^\mu k^\nu=\frac{(\rho+p_r)e^{-2\alpha}}{\tau(1+\kappa\rho)(1-\kappa p_r)}.
\label{eq:cc2}
\end{equation}
Now, for the auxiliary metric (\ref{eq:auxiliary_metric0}), it can be shown that
\begin{equation}
-R^t_t(\Gamma)+R^r_r(\Gamma)=-\frac{2}{H}e^{\nu-\Psi}\frac{d}{dr}\left[H'e^{-\nu-\Psi}\right],
\end{equation}
where a prime denotes a derivative with respect to $r$. Using (\ref{eq:relations}), the last equation can be rewritten as
\begin{equation}
-R^t_t(\Gamma)+R^r_r(\Gamma)=\sqrt{\frac{\tau(1-\kappa p_r)(1-\kappa p_\theta)}{(1+\kappa\rho)}}\left[-\frac{2}{r}e^{\alpha-\beta}\frac{d}{dr}\left(e^{-\alpha-\beta}\right)Y-\frac{2}{r}e^{-2\beta}Y'\right],
\label{eq:cc3}
\end{equation}
where $Y=\tau\sqrt{(1+\kappa\rho)(1-\kappa p_r)}H'$. Denoting the Ricci tensor of the physical metric $g_{\mu\nu}$ by $R_{\mu\nu}$, we obtain, for the physical metric (\ref{eq:physical_metric0}),
\begin{equation}
-R^t_t+R^r_r=-\frac{2}{r}e^{\alpha-\beta}\frac{d}{dr}\left(e^{-\alpha-\beta}\right),
\end{equation}
where the indices of $R_{\mu\nu}$ are raised by using the physical metric (\ref{eq:physical_metric0}). Therefore, along the family of radial null geodesics, we have
\begin{equation}
R_{\mu\nu}k^{\mu}k^{\nu}=(-R^t_t+R^r_r)e^{-2\alpha}=-\frac{2}{r}e^{-(\alpha+\beta)}\frac{d}{dr}\left(e^{-\alpha-\beta}\right).
\label{eq:cc4}
\end{equation}
Also, from Eqs. (\ref{eq:field_equation2}) and (\ref{eq:relations}), we obtain
\begin{equation}
R^t_t(\Gamma)=\frac{1}{\kappa}[1-\tau(1+\kappa \rho)].
\label{eq:cc5}
\end{equation}
Using Eqs. (\ref{eq:cc2}), (\ref{eq:cc3}), (\ref{eq:cc4}) and (\ref{eq:cc5}) in (\ref{eq:cc1}), we obtain, after some manipulations,
\begin{equation}
R_{\mu\nu}k^{\mu}k^{\nu}=\frac{(\rho+p_r)e^{-2\alpha}}{(1-\kappa p_r)X(r)}+\frac{2}{r}e^{-2(\alpha+\beta)}\frac{d}{dr}\log\left[\frac{(1+\kappa\rho)^{1/4}(1-\kappa p_r)^{1/4}}{(1-\kappa p_\theta)}X(r)\right],
\label{eq:NCC_EiBI}
\end{equation}
where
\begin{equation}
X(r)=\left[1+\frac{\kappa r}{4}\left(\frac{\rho'}{1+\kappa\rho}-\frac{p_r'}{1-\kappa p_r}\right)\right],
\end{equation}
and we have used the expression $H(r)=r/\sqrt{\tau(1-\kappa p_{\theta})}$. In the GR limit ($\kappa\to 0$), we obtain
\begin{equation}
\lim_{\kappa\to 0} R_{\mu\nu}k^{\mu}k^{\nu}=(\rho+p_r)e^{-2\alpha}.
\end{equation}
To satisfy the energy conditions, we must have $\rho+p_r\geq 0$ which, in turn, implies $R_{\mu\nu}k^{\mu}k^{\nu}\geq 0$ along the family of radial null geodesics in GR. Therefore, a violation/satisfaction of the NCC in GR means violations/satisfactions of different energy conditions. However, in EiBI gravity, the second term on the right-hand side of Eq. (\ref{eq:NCC_EiBI}), which vanishes in the limit $\kappa\to 0$, makes the difference between the NCC and the NEC. Therefore, in this gravity theory, the second term on the right-hand side of (\ref{eq:NCC_EiBI}) can lead to the violation of the NCC, which is required to maintain a wormhole, even though the NEC or any other energy conditions remain satisfied. In the next section, we show this explicitly by obtaining a class of wormhole solutions which violate the NCC but satisfy the NEC as well as all other energy conditions.

\section{Exact wormhole solutions satisfying all the energy conditions}
\label{sec:wormholes}

In the previous section, we have seen that the second term on the right-hand side of Eq. (\ref{eq:NCC_EiBI}), which vanishes in the GR limit $\kappa\to 0$, makes the difference between the NCC and the NEC. To see whether or not this second term alone can support wormholes without violating the energy conditions, we consider an energy-momentum tensor of the form $T^\mu_{\;\nu}={\rm diag}(-\rho,-\rho,p_\theta,p_\theta)$, such that $p_r=-\rho$ and the first term appearing on the right-hand side of Eq. (\ref{eq:NCC_EiBI}) vanishes. This type of energy-momentum can be interpreted as that due to an anisotropic fluid, or it can be obtained from a nonlinear electrodynamics action of the form
\begin{equation}
S_{M}=\frac{1}{8\pi}\int d^4x\sqrt{-g}\varphi(F),
\end{equation}
where $\varphi(F)$ is a function of the electromagnetic field invariant $F=-\frac{1}{2}F_{\mu\nu}F^{\mu\nu}$ and $F_{\mu\nu}=\partial_\mu A_\nu-\partial_\nu A_\mu$ is the electromagnetic field tensor. For the electrostatic case, the energy-momentum tensor obtained from the variation of the above action becomes \citep{olmo_universe},
\begin{equation}
T^{\mu}_{\;\nu}=\frac{1}{8\pi} {\rm diag}(\varphi-2F\varphi_F,\varphi-2F\varphi_F,\varphi,\varphi),
\end{equation}
where $\varphi_F$ is the derivative of $\varphi(F)$ with respect to $F$. Expressing the above energy-momentum tensor in the anisotropic fluid form, we have
\begin{equation}
\rho=-p_r =2F \varphi_F-\varphi, \hspace{0.3cm} p_\theta =\varphi,
\label{eq:nonlinear_electrodynamics}
\end{equation}
where we have set $8\pi=1$, as discussed in the previous section. For the above form of the energy-momentum tensor, the conservation equation $\nabla_\mu T^{\mu\nu}=0$ becomes
\begin{equation}
\rho'+\frac{2}{r}(\rho+p_\theta)=0.
\label{eq:conservation_eqn}
\end{equation}
In the subsequent calculations, we shall use the above conservation equation whenever $\rho'$ appears. Using the last equation, we obtain
\begin{equation}
X(r)=\frac{1-\kappa p_\theta}{1+\kappa \rho}.
\end{equation}
Therefore, Eq. (\ref{eq:NCC_EiBI}) becomes
\begin{equation}
R_{\mu\nu}k^{\mu}k^{\nu}=\frac{2\kappa}{r^2}\left(\frac{\rho+p_\theta}{1+\kappa \rho}\right) e^{-2(\alpha+\beta)}.
\label{eq:cc6}
\end{equation}
Now Eqs. (\ref{eq:cc4}) and (\ref{eq:cc6}) can be combined to obtain
\begin{equation}
\frac{d}{dr}\log\left(e^{\alpha+\beta}\right)=-\frac{d}{dr}\log\left(\sqrt{1+\kappa\rho}\right),
\end{equation}
where we have used the conservation equation (\ref{eq:conservation_eqn}). The integration of the last equation yields
\begin{equation}
e^{\alpha+\beta}=\frac{1}{\sqrt{1+\kappa\rho}}.
\label{eq:psi_0}
\end{equation}
Therefore, for an energy-momentum tensor of the form $T^\mu_{\;\nu}={\rm diag}(-\rho,-\rho,p_\theta,p_\theta)$ in EiBI gravity, we have, along a congruence of radial null geodesics,
\begin{equation}
R_{\mu\nu}k^{\mu}k^{\nu}=\frac{2\kappa}{r^2}\left(\rho+p_\theta\right).
\end{equation}
Note that, for the energy-momentum mentioned above, the necessary and sufficient conditions to satisfy all the energy conditions are $\rho\geq 0$, $p_\theta\geq 0$ and $\rho\geq \vert p_\theta\vert$. Therefore, we must have $\kappa<0$ for the violation of the NCC, and hence, to have wormholes without violating the energy conditions. The spacetime geometry of a spherically symmetric, static wormhole of the Morris-Thorne class is generically written as
\begin{equation}
ds^2=-e^{2\Phi(r)} dt^2+\frac{dr^2}{1-\frac{b(r)}{r}}+r^2(d\theta ^2+\sin ^2\theta d\phi ^2),
\label{eq:MT_wormhole1}
\end{equation}
where $\Phi(r)$ and $b(r)$ are, respectively, the redshift function and the wormhole shape function. The wormhole throat, where two different regions are connected, is given by $\left(1-\frac{b(r)}{r}\right)\Big\vert_{r_0}=0$, i.e., by $b(r_0)=r_0$, with $r_0$ being the radius of the throat. The redshift function $\Phi(r)$ is finite everywhere (from the throat to spatial infinity). Now, using (\ref{eq:psi_0}), we find that the physical metric (\ref{eq:physical_metric0}) becomes
\begin{equation}
ds_g^2=-e^{2\alpha(r)} dt^2+\frac{dr^2}{e^{2\alpha(r)}(1+\kappa\rho)}+r^2(d\theta ^2+\sin ^2\theta d\phi ^2).
\label{eq:MT_wormhole2}
\end{equation}
Comparing (\ref{eq:MT_wormhole1}) and (\ref{eq:MT_wormhole2}), we find that the above spacetime represents a wormhole, provided the throat radius $r_0$ is a solution of $(1+\kappa\rho)\vert_{r_0}=0$, and $\alpha(r)$ is finite from the throat to spatial infinity. This can also be seen from the fact that the expansion scalar
\begin{equation}
\hat{\theta}=\nabla_\mu k^\mu=\pm \frac{2}{r}e^{-(\alpha+\beta)}=\pm \frac{2}{r}\sqrt{1+\kappa\rho}
\end{equation}
of a congruence of radial null geodesics passing through the wormhole must vanish at the wormhole throat. Therefore, the wormhole throat radius $r_0$ must be a solution of $(1+\kappa\rho)\vert_{r_0}=0$.

To obtain exact wormhole solutions, we rewrite the physical and auxiliary metrics in the following forms:
\begin{equation}
ds_g^2=-\psi^2(r)f(r)dt^2+\frac{dr^2}{f(r)}+r^2\left(d\theta^2+\sin^2\theta d\phi^2\right),
\end{equation}
\begin{equation}
ds_q^2=-G^2(r)F(r)dt^2+\frac{dr^2}{F(r)}+H^2(r)\left(d\theta^2+\sin^2\theta d\phi^2\right).
\end{equation}
Comparing the above ansatze with Eqs. (\ref{eq:physical_metric0}) and (\ref{eq:auxiliary_metric0}), we find $e^{2\alpha}=\psi^2f$, $e^{2\beta}=\frac{1}{f}$, $e^{2\nu}=G^2F$ and $e^{2\Psi}=\frac{1}{F}$. Therefore, from Eqs. (\ref{eq:relations}) and (\ref{eq:psi_0}), we find that
\begin{equation}
f(r)=F(r)(1-\kappa p_\theta), \hspace{0.3cm} \psi(r)=\frac{1}{\sqrt{1+\kappa \rho}},
\label{eq:relation1}
\end{equation}
\begin{equation}
G(r)=\frac{1-\kappa p_\theta}{\sqrt{1+\kappa\rho}}, \hspace{0.3cm} H(r)=r\sqrt{1+\kappa \rho}.
\label{eq:relation2}
\end{equation}
Using Eq. (\ref{eq:relation2}) and the conservation equation (\ref{eq:conservation_eqn}), it can be shown that $H'=\frac{1-\kappa p_\theta}{\sqrt{1+\kappa\rho}}=G$. With $G=H'$, the $tt$ and $rr$ components of the field equation (\ref{eq:field_equation2}) become identical to each other. Therefore, we are left with two equations coming from the $tt$ (or $rr$) and the $\theta\theta$ component. The energy conservation equation (\ref{eq:conservation_eqn}) can be solved to obtain $\rho(r)$ for a given nonlinear electrodynamics model or equation of state between $\rho$ and $p_{\theta}$ of the anisotropic fluid. The $\theta\theta$ component of the field equation (\ref{eq:field_equation2}) can be solved to obtain $F(r)$. The other equation (i.e., the $tt$ or the $rr$ component of the field equation) will automatically be satisfied because of the energy conservation equation. The $\theta\theta$ component of the field is given by
\begin{equation}
2\frac{H''}{H}+\frac{H'^2}{H^2}+\frac{H'F'}{HF}-\frac{1}{H^2F}=\frac{1}{\kappa F}\left[\frac{1}{1+\kappa\rho}-1\right],
\end{equation}
which can be integrated to obtain
\begin{equation}
F= \frac{1}{HH'^2}\left[C_1+H-\frac{H^3}{3\kappa}+\frac{1}{\kappa}\int^r \frac{H^2H'}{1+\kappa\rho} dr\right],
\end{equation}
where $C_1$ is an integration constant. Therefore, we obtain
\begin{eqnarray}
f(r) &=& \frac{1-\kappa p_\theta}{HH'^2}\left[C_1+H-\frac{H^3}{3\kappa}+\frac{r^2H}{\kappa} -\frac{2}{\kappa}\int^r rH dr\right] \nonumber \\
&=&\frac{1+\kappa\rho}{1-\kappa p_{\theta}}\left[1+\frac{C_1}{r\sqrt{1+\kappa\rho}}-\frac{r^2}{3\kappa}(1+\kappa\rho)+\frac{r^2}{\kappa}-\frac{2}{\kappa r\sqrt{1+\kappa\rho}}\int^r r^2\sqrt{(1+\kappa\rho)} dr \right],
\label{eq:general_f}
\end{eqnarray}
where we have used $H'=\frac{1-\kappa p_\theta}{\sqrt{1+\kappa\rho}}$. Since in vacuum, the EiBI gravity reduces to vacuum GR, we must recover the Schwarzschild solution. This gives $C_1=-2M$, with $M$ being related to the mass. Therefore, we have obtained a complete general solution of the field equations. In the GR limit $\kappa\to 0$, $\psi=1$ and
\begin{eqnarray}
f(r)\big\vert_{\kappa\to 0}&=&1-\frac{2M}{r}-\frac{r^2}{3\kappa}(1+\kappa\rho)+\frac{r^2}{\kappa}-\frac{2}{\kappa r}\left(1-\frac{\kappa\rho}{2}\right)\int^r r^2\left(1+\frac{\kappa\rho}{2}\right) dr \nonumber\\
&=& 1-\frac{2M}{r}-\frac{1}{r}\int^r \rho r^2 dr,
\end{eqnarray}
which is the same as that in GR \citep{GR_NED}. For a Maxwell electric field ($p_\theta=\rho$), the integration of the energy conservation equation (\ref{eq:conservation_eqn}) gives $\rho=\frac{Q^2}{r^4}$, where $Q$ is an integration constant representing the charge. Putting this in the last equation, we obtain the metric function of the Reissner-Nordstrom spacetime.

The general solution (\ref{eq:general_f}) can represent both black holes and wormholes, depending on the signs and values of the different parameters. The black hole solutions are characterized by event horizons given by the roots of $f(r_H)=0$, with $r_H$ being the radius of an event horizon. However, in this work, we only analyze the wormhole solutions. As we have already shown, wormhole solutions are possible only when $\kappa<0$. The radius $r_0$ of a wormhole throat is given by $(1+\kappa\rho)\vert_{r_0}=0$. Other necessary conditions which have to be satisfied to construct a wormhole are the no-horizon condition and the flare-out condition at the throat. The metric function $\psi^2f$ must be nonzero, positive (no-horizon condition) and finite everywhere (from the throat to spatial infinity). However, because of the $(1+\kappa\rho)$ factor in the denominator of the terms containing the mass $M$ and the integration of (\ref{eq:general_f}), $\psi^2f$ diverges as $r\to r_0$. However, we can remove this divergence if we demand that, at the wormhole throat $r_0$,
\begin{equation}
M=\frac{1}{|\kappa|}\int^{r_0} r^2\sqrt{1+\kappa\rho} dr.
\label{eq:general_condition}
\end{equation}
In fact, the above condition not only removes the divergence in $\psi^2f$, it also removes the curvature divergences, thereby making the spacetime regular everywhere. This can be checked by finding the Ricci scalar $\mathcal{R}$. Expanding the metric functions around $r=r_0$ or using the l'H$\hat{\text{o}}$tal rule at $r=r_0$, we obtain
\begin{equation}
\mathcal{R}\big\vert_{r_0}=-\frac{(1-x)\kappa p_0'}{r(1-\kappa p_0)}+\frac{2x}{3r_0^2} \kappa p_0+\frac{2}{r_0^2}(2x-1),
\label{eq:ricci}
\end{equation}
where $x=\frac{r_0^2}{|\kappa|}$ and $p_0$ is the tangential pressure at the throat $r=r_0$, i.e., $p_0=p_\theta(r_0)$. Note that the Ricci scalar is finite at the throat. In terms of $f(r)$, the flare-out condition reads $\frac{f'}{2(1-f)^2}>0$ \citep{rajibul_2015}. At $r=r_0$, we have
\begin{equation}
f(r)\big|_{r_0}=0, \hspace{0.3cm} \psi^2(r)f(r)\big|_{r_0}=\frac{1}{1-\kappa p_0}(1-x)
\end{equation}
\begin{equation}
\frac{f'}{2(1-f)^2}\Big|_{r_0}=\frac{1}{r_0}(1-x).
\end{equation}
Note that, to satisfy the no-horizon condition as well as the flare-out condition at the throat, we must have $x<1$, i.e., $r_0<\sqrt{|\kappa|}$. Since, for $x<1$, $f=0$ and $f'>0$ at the throat, $f(r)$ does not have any zeroes at $r>r_0$. On the other hand, for $x>1$, it always possesses zeroes at $r>r_0$. Therefore, we always have a wormhole solution for $x<1$ and a regular black hole (or a wormhole whose throat is covered by an event horizon) solution for $x>1$. The critical value $x_c=1$ distinguishes the wormhole and black hole solutions.

\section{Some specific examples}
\label{sec:examples}
\subsection{Power-law Maxwell field}
For a power-law Maxwell electric field, $\varphi= F^{\beta}$. From Eq. (\ref{eq:nonlinear_electrodynamics}), we obtain
\begin{equation}
\rho=-p_r =(2\beta-1)F^\beta, \hspace{0.3cm} p_\theta =F^\beta=\alpha\rho,
\end{equation}
where $\alpha=\frac{1}{2\beta-1}$, i.e., $\beta=\frac{1+\alpha}{2\alpha}$. For $\alpha=1$, it represent the energy-momentum tensor of a Maxwell field. Wormhole solutions with the above type of energy-momentum tensor have already been obtained in \citep{rajibul_2015}. Here, we show that we can retrieve these solutions as a special case of the general solution (\ref{eq:general_f}). The energy conservation equation (\ref{eq:conservation_eqn}) can be integrated to obtain
\begin{equation}
\rho=\frac{C_0}{r^{2(\alpha+1)}},
\end{equation}
where $C_0$ is an integration constant. The integration in $f(r)$ can be performed to obtain
\begin{eqnarray*}
\int^r r^2\sqrt{1+\frac{\kappa C_0}{r^{2(\alpha+1)}}}dr &=& \frac{r^3}{3}\sqrt{1+\frac{\kappa C_0}{r^{2(\alpha+1)}}}+\frac{1}{3}(\alpha+1)\kappa C_0 I(r),
\end{eqnarray*}
where
\begin{eqnarray}
I(r)&=&\int^r \frac{1}{r^{2\alpha}\sqrt{1+\frac{\kappa C_0}{r^{2(\alpha+1)}}}} dr\label{eq:integration_I(r)}\\
&=& \left\{
  \begin{array}{lr}
    \frac{2}{3}\log\left[\left(\frac{r}{r_0}\right)^{\frac{3}{2}}+\sqrt{\left(\frac{r}{r_0}\right)^{3}\mp 1}\right] & : \alpha =\frac{1}{2}\\
  \frac{r^{1-2\alpha}}{1-2\alpha} {}_2F_1\left[\frac{1}{2},\frac{2\alpha-1}{2\alpha+2},\frac{4\alpha+1}{2\alpha+2};\pm\left(\frac{r_0}{r}\right)^{2\alpha+2} \right] & : \alpha \neq \frac{1}{2}
  \end{array}
\right. ,
\nonumber
\end{eqnarray}
where the upper and lower signs are for $\kappa<0$ and $\kappa>0$, respectively, and $r_0=(|\kappa| C_0)^{\frac{1}{2(\alpha+1)}}$. Therefore, we obtain
\begin{equation}
f(r)=\frac{1+\frac{\kappa C_0}{r^{2(\alpha+1)}}}{1-\frac{\alpha\kappa C_0}{r^{2(\alpha+1)}}}\left[1-\frac{2M}{r\sqrt{1+\frac{\kappa C_0}{r^{2(\alpha+1)}}}}-\frac{C_0}{3r^{2\alpha}}-\frac{2(\alpha+1)C_0}{3r\sqrt{1+\frac{\kappa C_0}{r^{2(\alpha+1)}}}}I(r) \right],
\nonumber
\end{equation}
which is the same as that obtained in \citep{rajibul_2015}. For $\kappa<0$, $r_0=(|\kappa| C_0)^{\frac{1}{2(\alpha+1)}}$ is the wormhole throat radius. Therefore, for $\kappa<0$, Eqs. (\ref{eq:general_condition}) and (\ref{eq:ricci}) become
\begin{eqnarray*}
M &=&-\frac{(\alpha+1)r_0^{2(\alpha+1)}}{3|\kappa|}I(r_0)\\
&=& \left\{
  \begin{array}{lr}
    0 & : \alpha =\frac{1}{2}\\
  \frac{(\alpha+1)r_0^3}{3(2\alpha-1)|\kappa|} \hspace{0.1cm} {}_2F_1\left[\frac{1}{2},\frac{2\alpha-1}{2\alpha+2},\frac{4\alpha+1}{2\alpha+2};1 \right] & : \alpha \neq \frac{1}{2}
  \end{array}
\right. ,
\end{eqnarray*}
\begin{equation}
\mathcal{R}\big|_{r_0}=-\frac{1}{r_0^2}\left[2(\alpha+1)-4\left(\frac{\alpha}{3}+1\right)x\right].
\nonumber
\end{equation}
The above results match with those obtained in \citep{rajibul_2015}. For the Maxwell electrodynamics $\alpha=1$ and $C_0=Q^2$, with $Q$ being the charge. In this Maxwell electrodynamics case, $f(r)$ becomes
\begin{equation}
f(r) = \left(\frac{1+\frac{\kappa Q^2}{r^4}}{1-\frac{\kappa Q^2}{r^4}}\right)\left[1-\frac{2M}{r\sqrt{1+\frac{\kappa Q^2}{r^4}}}-\frac{Q^2}{3r^2}+\frac{4Q^2}{3r^2\sqrt{1+\frac{\kappa Q^2}{r^4}}}\,{}_2F_1\left(\frac{1}{2},\frac{1}{4};\frac{5}{4};-\frac{\kappa Q^2}{r^4}\right)\right],
\end{equation}
where we have used $r_0^4=|\kappa|Q^2$ in $I(r)$. For $\kappa<0$, $r_0$ is the throat radius.

\subsection{Born-Infeld electrodynamics}
For a static electric field in Born-Infeld electrodynamics, $\varphi$ is given by
\begin{equation}
\varphi(F)=2b^2\left(1-\sqrt{1-\frac{F}{b^2}}\right),
\label{eq:born_infeld_electrodynamics}
\end{equation}
where $b$ is the Born-Infeld electrodynamics parameter. In the limit $b^2\to \infty$, it reduces to Maxwell electrodynamics. Black hole solutions in EiBI gravity coupled to the above Born-Infeld electrodynamics have been obtained in \citep{soumya_2015}. Here, we highlight the wormhole solutions supported by the above nonlinear electrodynamics. From Eqs. (\ref{eq:nonlinear_electrodynamics}) and (\ref{eq:born_infeld_electrodynamics}), it can be shown that
\begin{equation}
p_\theta=\frac{\rho}{1+\frac{\rho}{2b^2}},
\label{eq:BI_pressure}
\end{equation}
which can be used to integrate the conservation equation (\ref{eq:conservation_eqn}) to obtain
\begin{equation}
\rho=2b^2\left(\sqrt{1+\frac{Q^2}{b^2r^4}}-1\right),
\label{eq:BI_energy}
\end{equation}
where $Q$ is an integration constant representing the charge. In this case, however, it is difficult to perform the integration in $f(r)$ analytically in the Schwarzschild gauge. To perform the integration analytically, we consider the coordinate transformation
\begin{eqnarray}
H(r)&=&r\sqrt{1+\kappa\rho}=\bar{r}(r) \nonumber\\
\Rightarrow \bar{r}(r)&=&r\sqrt{1+2\kappa b^2\left(\sqrt{1+\frac{Q^2}{b^2r^4}}-1\right)}
\label{eq:coordinate_transformation}
\end{eqnarray}
and obtain
\begin{equation}
\int^r r^2\sqrt{(1+\kappa\rho)} dr=\frac{1}{2}r^2\bar{r}-\frac{1}{2}\int^r r^2\bar{r}' dr=\frac{1}{2}r^2\bar{r}-\frac{1}{2}\int^{\bar{r}} \frac{\bar{r}^2}{1+\kappa\rho} d\bar{r}.
\label{eq:BI_integration_1}
\end{equation}
Now putting $r=\bar{r}/\sqrt{1+\kappa\rho}$ in (\ref{eq:coordinate_transformation}), we obtain
\begin{equation}
\frac{1}{1+\kappa\rho}=\frac{1-2\kappa b^2\left(1+\sqrt{1+\frac{Q^2}{b^2\bar{r}^4}-\frac{4\kappa Q^2}{\bar{r}^4}}\right)}{1-4\kappa b^2}.
\end{equation}
Using the above expression and defining $4\kappa b^2=\alpha$, the integration on the right-hand side of (\ref{eq:BI_integration_1}) becomes
\begin{eqnarray}
\int^{\bar{r}} \frac{\bar{r}^2}{1+\kappa\rho} d\bar{r}&=&\frac{2-\alpha}{6(1-\alpha)}\bar{r}^3-\frac{\alpha}{2(1-\alpha)}\int^{\bar{r}} \bar{r}^2\sqrt{1+\frac{4\kappa Q^2(1-\alpha)}{\alpha\bar{r}^4}}d\bar{r}\nonumber\\
&=&\frac{2-\alpha}{6(1-\alpha)}\bar{r}^3-\frac{\alpha}{6(1-\alpha)}\bar{r}^3\sqrt{1+\frac{4\kappa Q^2(1-\alpha)}{\alpha\bar{r}^4}}\nonumber\\
& & -\frac{4}{3}\kappa Q^2\int^{\bar{r}}\frac{d\bar{r}}{\bar{r}^2\sqrt{1+\frac{4\kappa Q^2(1-\alpha)}{\alpha\bar{r}^4}}}.
\label{eq:BI_integration_2}
\end{eqnarray}
Note that the integration on the right-hand side of (\ref{eq:BI_integration_2}) is similar to that in Eq. (\ref{eq:integration_I(r)}) (with $\alpha=1$ there). This gives
\begin{equation}
\int^{\bar{r}}\frac{d\bar{r}}{\bar{r}^2\sqrt{1+\frac{4\kappa Q^2(1-\alpha)}{\alpha\bar{r}^4}}}=-\frac{1}{\bar{r}} {\;}_2F_1\left[\frac{1}{2},\frac{1}{4},\frac{5}{4};-\frac{4\kappa Q^2(1-\alpha)}{\alpha\bar{r}^4} \right].
\label{eq:BI_integration_3}
\end{equation}
Combining (\ref{eq:BI_integration_1}), (\ref{eq:BI_integration_2}) and (\ref{eq:BI_integration_3}), we obtain
\begin{eqnarray}
\int^r r^2\sqrt{(1+\kappa\rho)} dr &=& \frac{\bar{r}^3}{6(1-\alpha)}\left(2-\alpha-\alpha\sqrt{1+\frac{4\kappa Q^2(1-\alpha)}{\alpha\bar{r}^4}}\right)\nonumber\\
& & -\frac{2\kappa Q^2}{3\bar{r}} {\;}_2F_1\left[\frac{1}{2},\frac{1}{4},\frac{5}{4};-\frac{4\kappa Q^2(1-\alpha)}{\alpha\bar{r}^4} \right].
\label{eq:BI_integration_4}
\end{eqnarray}
So, finally, we obtain
\begin{eqnarray}
f(\bar{r}) &=& \frac{1+\kappa\rho}{1-\kappa p_\theta}\left[1-\frac{2M}{\bar{r}}+\frac{\alpha\bar{r}^2}{6\kappa(1-\alpha)}\left(1-\sqrt{1+\frac{4\kappa Q^2(1-\alpha)}{\alpha\bar{r}^4}}\right)\right. \nonumber\\
& & \left. +\frac{4Q^2}{3\bar{r}^2}{\;}_2F_1\left[\frac{1}{2},\frac{1}{4},\frac{5}{4};-\frac{4\kappa Q^2(1-\alpha)}{\alpha\bar{r}^4} \right]\right],
\end{eqnarray}
which is the same as that obtained in \cite{soumya_2015} for $-\infty<\alpha<1$. In the above expression, $p_\theta$, $\rho$ and $\bar{r}(r)$ are, respectively, given by (\ref{eq:BI_pressure}), (\ref{eq:BI_energy}) and (\ref{eq:coordinate_transformation}). The above solution represents a wormhole for $\kappa<0$, i.e., for $\alpha<0$. The wormhole throat radius $r_0$ is given by $(1+\kappa\rho)|_{r_0}=0$. This gives $\bar{r}(r_0)=0$ or $r_0=\left[4|\kappa|^2b^2Q^2/(1+4|\kappa|b^2)\right]^{1/4}$. In the Maxwell electrodynamics limit $b^2\to \infty$, the throat radius becomes $r_0=(|\kappa|Q^2)^{1/4}$ which is the same as that obtained in the previous subsection.

\subsection{Anisotropic fluid with $p_\theta=\rho(1+\kappa\rho)$}
As the third example, we consider an anisotropic fluid with the equation of state $p_\theta=\rho(1+\kappa\rho)$. The energy conservation equation (\ref{eq:conservation_eqn}) can be integrated to obtain
\begin{equation}
\rho=\frac{C_0}{r^4}\frac{1}{1-\frac{\kappa C_0}{2r^4}},
\nonumber
\end{equation}
where $C_0$ is an integration constant. Putting $r=\frac{1}{z}$, the integration in $f(r)$ can be performed using MATHEMATICA. We obtain
\begin{eqnarray*}
\int^r r^2\sqrt{(1+\kappa\rho)} dr&=&-\int^z \frac{\sqrt{1+\frac{\kappa C_0}{2}z^4}}{\sqrt{1-\frac{\kappa C_0}{2}z^4}}\frac{dz}{z^4}\\
&=& \frac{1}{3z^3}\sqrt{1-\frac{\kappa^2 C_0^2}{4}z^8}-\frac{\kappa C_0}{2}z {\;}_2F_1\left[\frac{1}{2},\frac{1}{8},\frac{9}{8};\frac{\kappa^2C_0^2}{4}z^8 \right]\\
&& +\frac{\kappa^2 C_0^2}{4}\frac{z^5}{15} {\;}_2F_1\left[\frac{1}{2},\frac{5}{8},\frac{13}{8};\frac{\kappa^2 C_0^2}{4}z^8 \right].
\end{eqnarray*}
Therefore, we obtain, after some manipulations,
\begin{eqnarray}
f(r) &=& \frac{1+\kappa\rho}{1-\kappa p_{\theta}}\left[1-\sqrt{\frac{1-\frac{\kappa C_0}{2r^4}}{1+\frac{\kappa C_0}{2r^4}}}\left(\frac{2M}{r}-\frac{C_0}{r^2}{\;}_2F_1\left[\frac{1}{2},\frac{1}{8},\frac{9}{8};\frac{\kappa^2C_0^2}{4r^8} \right]\right. \right. \nonumber\\
& & \left. \left. +\frac{\kappa C_0^2}{30r^6}{\;}_2F_1\left[\frac{1}{2},\frac{5}{8},\frac{13}{8};\frac{\kappa^2 C_0^2}{4r^8} \right]\right) -\frac{\kappa C_0^2}{6r^6}\right].
\end{eqnarray}
For $\kappa<0$, the wormhole throat radius is given by $(1+\kappa\rho)\vert_{r_0}=0$ which gives $r_0=\left(\frac{|\kappa| C_0}{2}\right)^{1/4}$. In this case, Eqs. (\ref{eq:general_condition}) and (\ref{eq:ricci}) become
\begin{equation}
M=\frac{r_0^3}{|\kappa|} {\;}_2F_1\left[\frac{1}{2},\frac{1}{8},\frac{9}{8};1 \right]+\frac{r_0^3}{15|\kappa|} {\;}_2F_1\left[\frac{1}{2},\frac{5}{8},\frac{13}{8};1 \right],
\end{equation}
\begin{equation}
\mathcal{R}\big\vert_{r_0}=\frac{2x}{r_0^2}.
\end{equation}
Note that, for $\kappa<0$, $p_\theta$ vanishes at the throat and approaches $\rho$ asymptotically. Therefore, $0\leq p_{\theta}\leq \rho$ always, implying that all the energy conditions are satisfied.

\section{Conclusion}
\label{sec:conclusion}
In this work, we have established a relationship between the NCC and the NEC in EiBI gravity. We have shown that, in contrast to GR, in EiBI gravity, a violation of the NCC does not necessarily lead to violations of the various energy conditions, thereby implying that wormholes can be supported by nonexotic matter in this gravity theory. Subsequently, we have obtained exact solutions of the field equations in EiBI gravity coupled to arbitrary nonlinear electrodynamics and anisotropic fluids having energy-momentum of the form $T^\mu_{\;\nu}={\rm diag}(-\rho,-\rho,p_\theta,p_\theta)$. Depending on the signs and values of different parameters, the general solutions can represent both black holes and wormholes. In this work, we have analyzed the wormhole solutions. We have found that the EiBI theory parameter $\kappa$ must be negative so that the wormholes are supported by matter which satisfies all the energy conditions, even though the NCC is violated. As special cases of our general solutions, we have obtained several specific wormhole solutions by considering Maxwell, power-law, Born-Infeld electrodynamics models and an anisotropic fluid having energy-momentum of the form $T^\mu_{\;\nu}={\rm diag}(-\rho,-\rho,\rho(1+\kappa\rho),\rho(1+\kappa\rho))$. Currently, we are studying the black hole aspects of these solutions and hope to report our results in the future.

\section*{Acknowledgments}
The author acknowledges the Council of Scientific and Industrial Research, India under whose fellowship program a part of the work was done at IIT Kharagpur. He also acknowledges Sayan Kar for a careful reading of the manuscript.

\end{document}